# Third order nonlinear study of ZnO nano particles under femto-second laser illumination

A. Srinivasa Rao[1,3*], Gayathri Sethuraman[2], Oriparambil Sivaraman Nirmal Ghosh[2], Annamraju Kasi Viswanath[2] and Alok Sharan[1]

[1]Department of Physics, Pondicherry University, Puducherry-605014, India
[2]Nanophotonics and Nanoelectronics Research Laboratory, Centre for Nanoscience and Technology, Pondicherry University, Puducherry-605014, India.
[3]Photonic Sciences Lab, Physical Research Laboratory, Ahmedabad-380009, India
[*] asvrao@prl.res.in, sri.jsp7@gmail.com

**Abstract**

ZnO nano particles are synthesized at different percentages of Ag dopant by facile precipitation method. Two photon absorption (2PA) cross-section and nonlinear refractive index of ZnO nano particles are studied with 130 femto-second laser pulses at a repletion rate of 1 kHz with central wavelength of 532 nm by z-scan technique. The 2PA cross-section of ZnO nano particles are obtained by theoretical calculations, using rate equation approach. Both absorption cross-section and nonlinear refractive index increased with increasing the Ag dopant percentage. We have observed an increase of order of $10^4$ in the nonlinear refractive index in the presence of femto-second laser excitation with reference to nano-second laser excitation.

## 1. Introduction

Recent development in the synthesis of nano-materials created a band-gap engineered metal oxide nano-structures [1-3]. Wide variety of nano-metallo-oxides such as ZnO, $TiO_2$ and $SnO_2$ have been used for various applications in the photo-catalysis, electronics, photonics and magnetism [4-9]. ZnO has a wide band gap of 3.37 eV and large exciton binding energy around 60 meV [10]. On the other side, nonlinear absorptive materials have the two kinds of behavior in their absorption nonlinear process: saturable absorbers [11,12] and optical limiters [13,14]. The high electron mobility nature of ZnO can be used as an optical limiter under multi-photon absorption. The band-gap engineering approach to modify the crystal defects and electronic structure of ZnO is used to tailor make materials for desired optical properties. Synthesis of Ag doped ZnO nanostructures with modified surface properties facilitates the interfacial charge transfer process for effective utilization of conduction band electrons which can enhance the nonlinear optical properties.

To synthesize pristine ZnO nano-particles at different Ag percentage, we adopted the facile precipitation method [15]. The ZnO-Ag is suspended in ethanol solution prepared at molar concentration of $1.8 \times 10^{-4}$ mol/cm$^3$. The sizes of nano-particles prepared are around 30 nm and the full details about sample characterization can be found in Ref. 16. The intercept on the wavelength axis corresponds to band-gap at 380 nm. The particle size decreases with increasing the dopant percentage. Also the band gap of ZnO nano particles increased with dopant percentage. ZnO material is a widely used material for different applications involving both linear and nonlinear optical region, including multi-photon absorption. Nonlinear optical properties of ZnO have been well studied by cw [17], nano [18-23], pico [24] and femto-second pulsed lasers [25] with different dopant [26-28]. Nonlinear properties of ZnO are larger for the form of nano particles as comparable with its solid form. For a diverse range of applications, it has been cast in different forms like thin films [23, 29], nanodots [30], nano rods [31], nano tubes [32]. Ag was doped in ZnO at 1, 2 and 3 percentages to study the effect of dopant on the optical properties of ZnO.

## 2. Theory

In the presence of femto-second laser excitation, to study the population redistribution in the ZnO nano particles, we have used rate equation approach, which is simplest and efficient model [33-34]. Molecules pumped from ground to excited state at pumping rate $W_{01}=(I(t)/h\nu)\sigma_{01}$. The absorption cross-section from ground to excited state in the presence of 2PA is given as $\sigma_{01}=\sigma_{2PA}(I/h\nu)$. Let $m_0$ and $m_1$ be the fractional population occupying in the valance and the conduction band ($m_0=M_0/M$, $m_1=M_1/M$ and $M_0+M_1=M$, here $M$ is the total number of molecules/atoms interacting with the laser beam) and $I(t)$ is the intensity of the incident pulses. The rate equations will have the form as shown in the Eq. 1.

$$m_0 + m_1 = 1 \qquad (1a)$$

$$\frac{dm_0}{dt} = -\frac{\sigma_{01}I(t)}{h\nu}(m_0 - m_1) \qquad (1b)$$

$$\frac{dm_1}{dt} = \frac{\sigma_{01} I(t)}{h\nu}(m_0 - m_1) \tag{1c}$$

The integration over the pulse-width gives the fractional population in the ground and excited states are

$$m_0 = \frac{1}{2}\left[1 + \exp\left(-2\frac{I_0 \sigma_{01}}{h\nu}\int_{-\infty}^{+\infty} I(t)dt\right)\right] \tag{2a}$$

$$m_1 = \frac{1}{2}\left[1 - \exp\left(-2\frac{I_0 \sigma_{01}}{h\nu}\int_{-\infty}^{+\infty} I(t)dt\right)\right] \tag{2b}$$

The 2PA coefficient in terms of absorption cross-section is $\alpha_2 = M(m_0-m_1)\sigma_{2PA}/(h\nu)$ and the transmittance through the material is [35]

$$T = \frac{I_p e^{-\alpha L}}{q_0 \tau_{FWHM}}\sqrt{\frac{4\ln 2}{\pi}}\int_{-\infty}^{\infty}\ln\left|1 + q_0 \exp(-4\ln 2 t^2/\tau_{FWHM}^2)\right|dt \tag{3}$$

$$q_0 = \alpha_2 I_p L_{eff}; L_{eff} = \frac{1-e^{-\alpha L}}{\alpha}$$

Here $\alpha$ is the linear absorption coefficient, $L$ is the thickness of the sample and $I_p$ is the peak power of the pulse. In the presence of femto second laser excitation due to its high peak power the population in the ground state depleted even though it is 2PA processed. Therefore, it is necessary to take the contribution of excited states in the refractive index change. The change in the refractive index in the presence of two levels, we obtained as [36]

$$\Delta n = \frac{n_2 I}{\left(1 + 2I/I_{S01}\right)} \tag{4}$$

Here $n_2$ is a nonlinear refractive index. The phase distortion in the laser beam is $\Delta\Phi = (2\pi/\lambda)\Delta n L$. The z-scan closed aperture transmittance in terms of phase distortion given by [37]

$$T(z) = \frac{1}{1 - \frac{4x\Delta\Phi_0}{(1+x^2)^2} + \frac{4x\Delta\Phi_0^2}{(1+x^2)^3}} \tag{5}$$

where, $x = \sqrt{1 + \frac{z^2}{z_0^2}}$

## 3. Experimental results and discussions

To study the 2PA of ZnO nano particles in the presence of Ag dopant with different percentage concentration we have used z scan technique [38,39]. Raleigh length ($z_R$) and beam waist ($\omega_0$) of z-scan beam profile are 8 mm and 40 μm. In the experimental process, care has been taken to prevent the artifacts of experimental setup and data analysis [40]. We have excited with 532 nm central wavelength femto-second laser with 130 fs pulse width at 1 kHz repetition rate. The average power used for excitation is 3 mW to study the 2PA in ZnO nano-particles in the ethanol solution. Open and closed aperture z-scan of ZnO with different percentage of Ag dopant has been plotted in the Fig. 1. In the open aperture data, solid lines are fits corresponds to transmittance curve obtained from solutions of rate equations (Eq. 3). The closed aperture data is analyzed using theoretical Eq.5. From theoretical fits of the experimental data, we have obtain the 2PA absorption cross-section and nonlinear refractive index and these values are tabulated in Table 1. The 2PA cross-section and nonlinear refractive index values of Pure ZnO nano particles are $0.9 \times 10^{-49}$ cm$^4$s/photon and $5.0 \times 10^{-17}$ cm$^2$/W respectively. Using z-scan, Litty Irimpan et al., carried out the study of 2PA in ZnO at different Ag dopant percentages by exciting with ns laser at 532 nm wavelength [41]. Similar to the nanosecond laser excitation, the absorption nonlinearity was found to increase with Ag dopant in the femto-second laser excitation due increase in the absorption cross-section. The 2PA cross-section value increased to $1.1 \times 10^{-49}$ cm$^4$s/photon for 1% Ag dopant and $1.4 \times 10^{-49}$

cm$^4$s/photon for 3% Ag dopant. While, Ag nano particles shows a negative nonlinear refractive index in the nano-second time-scale, whereas in femto-second time-scale, we find it to demonstrate positive refractive nonlinearity. The nonlinearity in the refractive index increased with the increasing dopant percentage. At 1% and 3% Ag dopant, the ZnO nano particles nonlinear refractive index increases to 5.2×10$^{-17}$ cm$^2$/W and 6.1×10$^{-17}$ cm$^2$/W respectively without changing the sign of nonlinear refractive index. We have seen an increase of order of 10$^4$ in the magnitude of nonlinear refractive index due to femto-second laser excitation as compare with nano-second laser study. Thus, the absorption nonlinear as well as refractive nonlinear properties of ZnO can be controlled as a function of Ag dopant. The dependence of linear absorption coefficient on dopant percentage provides the advantage of control tuning of linear absorption. Using linear absorption control, we can set the initial threshold intensity of optical limiters for required specifications.

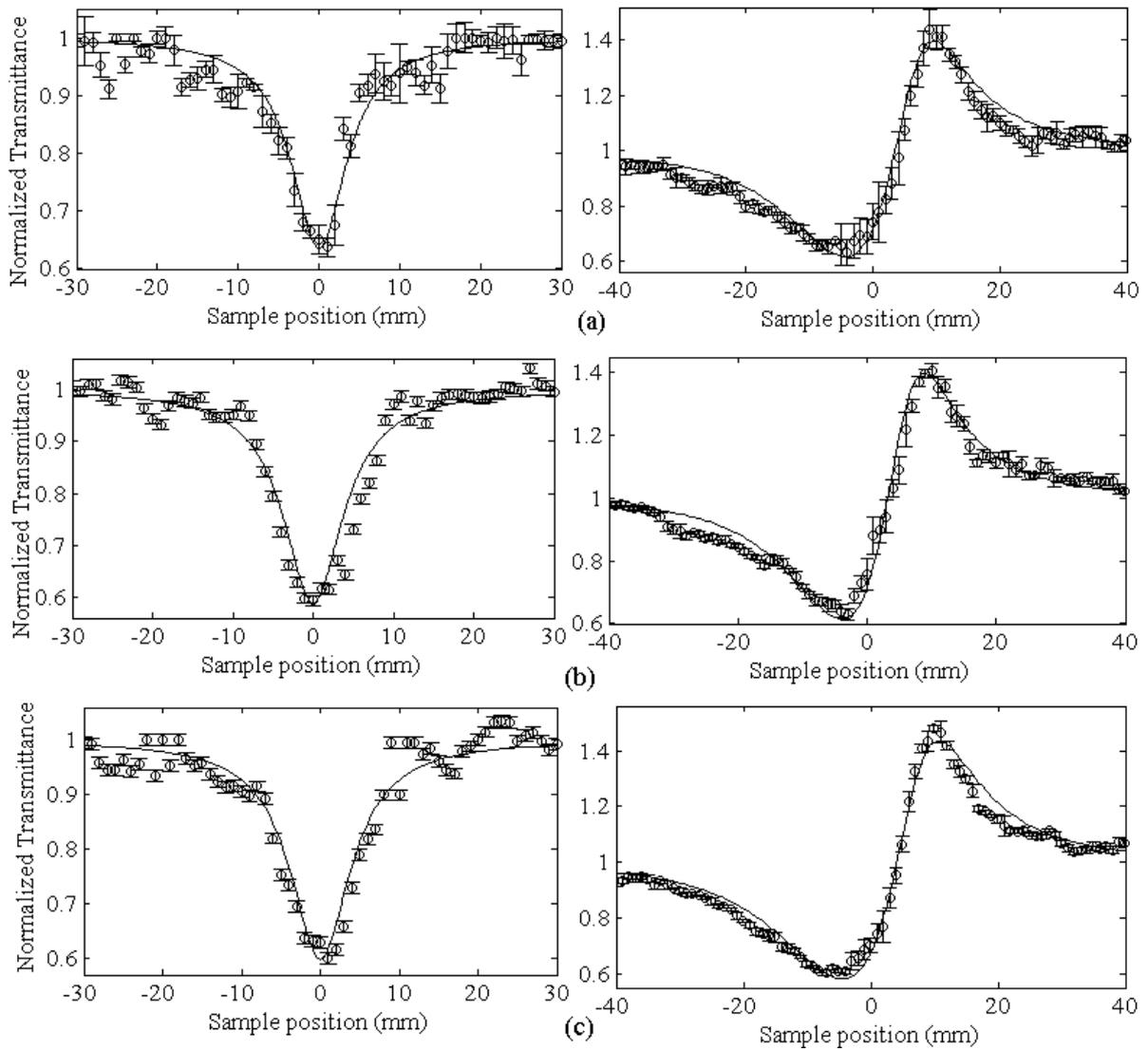

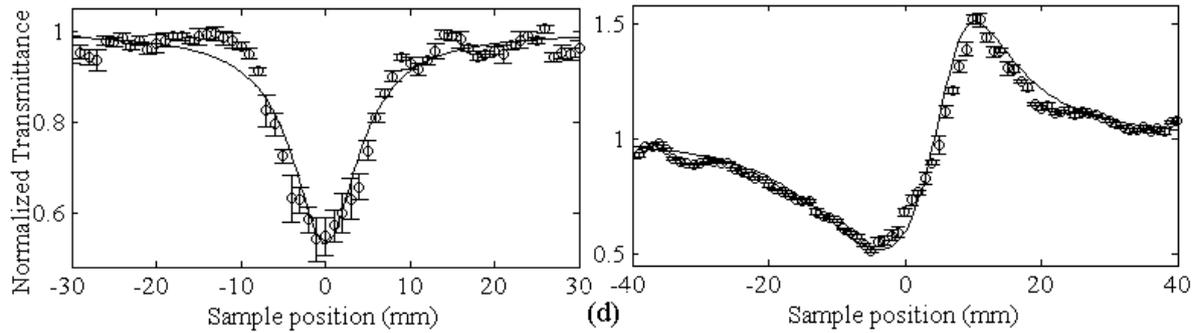

Fig. 1 Open and closed aperture z-scans of (a) pure ZnO, (b) ZnO with 1% Ag dopant, (c) ZnO with 2% Ag dopant and (d) ZnO with 3% Ag dopant.

Table 1 z-scan data of ZnO at different Ag dopant percentage

| Sample | Linear absorption $\alpha$ (1/cm) | 2PA cross-section $\sigma_{2PA}$ ($cm^4$ s/photon) | Nonlinear refractive index $n_2$ ($cm^2$/W) |
|---|---|---|---|
| ZnO | 3.8 | $0.9 \times 10^{-49}$ | $5.0 \times 10^{-17}$ |
| ZnOAg (1%) | 5.1 | $1.1 \times 10^{-49}$ | $5.2 \times 10^{-17}$ |
| ZnOAg (2%) | 5.4 | $1.1 \times 10^{-49}$ | $5.5 \times 10^{-17}$ |
| ZnOAg (3%) | 6.0 | $1.4 \times 10^{-49}$ | $6.1 \times 10^{-17}$ |

## 5. Conclusion

ZnO nano particles were locally synthesized at different Ag percentage by the facile precipitation method. By z-scan technique, while 2PA cross-section obtained by open aperture scan, closed aperture scan was used to obtain the nonlinear refractive index at large phase distortion. We have derived equations for population in the energy levels by rate equation approach for femto-second laser excitation and then obtained the transmittance as a function of peak intensity and absorption cross-section. In the presence of 130 femto-second laser excitation at 1 kHz pulse repletion rate, we have noticed that 2PA cross-section at 532 nm central wavelength was increased with increasing the Ag dopant percentage. We have obtained equation for change in the refractive index in the presence of two levels resonant and used to obtain the nonlinear refractive index coefficient under large phase distortion. The value of nonlinear refractive index was increased with the dopant. The Ag dopant controlled ZnO can be used as a good optical limiter.

**Acknowledgement:** Authors would like to thank CIF Pondicherry University for providing femto-second laser facilities.